
\magnification 1200

\centerline{\bf Non - Topological Solitons in non-minimally coupled}
\centerline{\bf Scalar fields: Theory and consequences}

\centerline{}

\centerline{}
\centerline{}
\centerline{Daksh Lohiya*}
\centerline{}
\centerline{\it Inter University Center for Astronomy and Astrophysics}
\centerline{\it [IUCAA], Postbag No.4, Ganeshkhind,}
\centerline{\it Pune 411 004, INDIA}

\vskip 2cm

\centerline{\bf Abstract}

	In theories of gravitation in which dimensional parameters are dynamically
induced, one can have non - topological - soliton solutions. This article
reviews related topics connected with such solutions.

	The existence of such solutions in curved spacetime
can give rise to
halos of gravity (g-) balls with gravitational ``constant" having different
values inside and outside the ball. Such g - balls can have quite interesting
bearing on the dark matter problem over galactic and cluster scales. We
describe
the origin of such solutions. We
speculate on related problems in Cosmology. Such objects would naturally
occur in a large class of induced gravity models in which we have scalar fields
non minimally coupled to the scalar curvature.

\vskip 2cm
IMPERIAL/TP/92-93/57. (*) Permanent address: Department of Physics, University
of Delhi, Delhi 110 007, INDIA.

\vfil\eject

Introduction: Non - topological soliton solutions are non trivial finite energy
solutions to the field equations which are stable. By a simple scaling
argument, it was demonstrated by Derrick [1964] that for a simple scalar field
theory, as also for a multicomponent theory, no static,
stable solutions can exist. In
curved spacetime it is possible to get aroung the Derrick constraint. First of
all the scaling argument would not work for arbitrary metric factors present in
the volume integral expression for the energy. Secondly it is possible to give
up the positivity of the effective potential in the theory as well as the
requirement that the integral over all space of the effective potential be
bounded. Further in curved spacetime the boundary conditions on a scalar field
for
a soliton solution can be chosen so that the effective potential tends to a
non - vanishing value at large distances outside the soliton.
In a flat spacetime
this is not permissible as it implies an infinite energy for the configuration.
However in curved spacetime, this corresponds to a non - vanishing cosmological
constant. The asymtotic Poincare invariance is lost in such a case. For an
asymtotic de Sitter or an anti de Sitter spacetime, the Poincare operator
$P_\mu^2$ is no longer a Casimir operator. Instead the 5 - dimensional
``rotation" generators $J_{ab}$ [defined as the generators of the O(4,1) or
the O(3,2) algebra] have the property that the $J_{04}$ becomes $P_0$ in the
limit $\Lambda \longrightarrow 0$. This property can be consistently used to
define the flux integral at infinity [Abbot and Deser 1982]
- thereby justifying non trivial
boundary conditions in curved spacetime.
In this article we shall outline reasons  that lead us to believe
in the existence of spherically symmetric soliton
solutions not only when there are conserved charges in a multicomponent
scalar field theory but even in the case of a simple
single component scalar field non - minimally coupled to the scalar curvature.
The solution represents a spherically symmetric region with an effective
gravitational constant having a value differing from the corresponding value
outside the region. Such a positive energy soliton, if it exists, would be both
classically and quantum mechanically stable.

	We have been exploring possible applications of such solutions. With
this objective in mind, section I considers the conventional formulation
of Einstein's theory. We start with an outline of the achievements and problems
in the conventional classical theory of gravity and describe the problems
one faces in quantization of the theory. Briefly recalling arguments connected
with a semiclassical treatment of gravity, we describe the induced gravity
programme in some detail. This is because the programme gives calculable value
of an induced gravitational and cosmological constants. However in a whole
class
of these theories, naive spectral decomposition of their expressions imply that
there is nothing in the theory that fixes the sign of the gravitational
constant.
What we have in mind is putting in possible soliton solutions in this program
as
a possible application of such solutions. Whatever be the sign and magnitude
of the induced gravitational constant, one could ensure the right sign and
magnitude of the constant in the presence of the soliton solution on account of
the non - minimal coupling of the scalar field to the curvature scalar.

	In section II we outline the various attempts made to construct a
dynamic classical theory of gravity by using scalar fields non - minimally
coupled
to the scalar curvature. Typically in such approaches, the gravitational
constant
gets related to the bulk properties of the universe. If the effective potential
of a scalar field has a non - degenerate, non - trivial minima, giving rise to
spontaneous symmetry breaking, the gravitational constant gets related to the
value of the field minimising the effective potential. We end the section by
recalling the necessity of a non minimal coupling for a scalar field in order
to have finite [cut off independent] matrix elements of the physical stress
energy tensor in a renormalised perturbation theory.

	In section III we establish suggestive arguments pointing to the existence,
stability and properties of soliton solutions in classes of
non minimally coupled scalar
field theories. In conclusion we discuss the possible applications of these
solutions.

Section I. In the conventional formulation of general relativity,
gravity is introduced
by the following ansatz: (i) re - write matter action in a generally covariant
form, and (ii) add to it the Einstein - Hilbert gravitational action:
$$ S = (16\pi G)^{-1}\int d^4 x \sqrt{-g} [R - 2\Lambda] \eqno{(1.1)} $$
[here G is the Newtonian gravitational constant, R the scalar curvature of the
spacetime manifold and $\Lambda ( \approx 0)$ the cosmological constant.] The
classical structure of this ansatz has been studied in considerable detail. The
theory supports spherically symmetric Einstein metric solutions [$R_{\mu\nu} =
\Lambda g_{\mu\nu}$]
which account for the three classical tests of general relativity.

	Strictly
speaking, though, these solutions do not single out Einstein's theory as
they are not unique to the second order Einstein - Hilbert theory. Any Einstein
metric solution [solution to $R_{\mu\nu} = \Lambda g_{\mu\nu}$] is also a
solution to field equations following from a fourth order theory of gravity:
with the action given by the spacetime integral of the Weyl tensor - squared.

	As regards the large scale problems of the universe, the theory
provides a framework
within which one could address a variety of problems in Cosmology. Of these,
particularly heralded as ``success stories" (modulo the horizon problem)
are the primodial light element
synthesis and the relic [micro - wave background] radiation.
Further, the above successes merely require a large
expansion of the conformal scale factor in cosmology to account for the relic
radiation and, the light element synthesis is not very
sensitive to the details of the standard big - bang model. When applied to
structure formation, the optimism that one can understand large scale structure
is quenched by the requirement that: (a) more than $90\%$ of the gravitating
content of the universe is non - luminous and in [as yet undecided] unknown
form. Attempts to account for large scale structure in the Universe by a
suitable combination of hot and cold dark matter has met with a doubtful
amount of success [see eg. Padmanabhan 1993, for review]. Fixing the total
amount of dark matter by reference to the closure density of the universe
and fixing the hot component by the requirement to form very large scale
structures, it is not clear whether one is left with
sufficient cold dark matter to account for
enough power to accomodate the smaller [gallactic and cluster] scale
structures.  Further, (b) the proximity of the density of the universe
to the closure
density poses yet another problem viz. the flatness problem. Related to this
problem is the difficulty to obtain a homogenous and isotropic universe
dynamically [see eg. MacCallum 1979]. The FRW metric is a very special metric
with a measure zero in the space of isotropic solutions.

	 On the quantum
theoretical side we have further problems. There is no viable
quantum theory of gravity within the framework of field theory. If we treat
the Einstein - Hilbert action as a fundamental quantum action, we get a
non - renormalizable quantum field theory.
One ``cure" to this would be to rule
that the gravitational field has a distinguished place in Physics and must not
be subject to quantisation. In a semi - classical theory of gravity, the
Einstein tensor is related to the expectation value of the quantised stress -
energy tensor of matter fields:
$$ G_{\mu\nu} = \kappa <T_{\mu\nu}>$$
Such a theory poses interesting possibilities [Kibble 1982]. One can envisage
a two - outcome quantum decision making arrangement eg.
a Stern - Gerlach apparatus
splitting a beam of spin - 1/2 atoms [polarised in the x - direction] into two
- according to the value of the y - component of their spins. Two detectors
arranged to intercept the two portions of the split beams are coupled to a
mechanical arrangement which drive a large mass M towards the detector which
clicks. A test particle placed along the line of the original beam would follow
the mass M. The beam, the detectors and the position of M form a two state
quantum system. Consider one atom passing through the system. If the
gravitational
field is quantised then superposition principle holds and the test particle
would follow M as a detector clicks. However if semi - classical theory holds,
the theory is inherently non linear and the
superposition principle would not hold. The
gravitational field acting on the test particle would be given by the
expectation
value of the stress tensor: i.e. it would get equal contributions from
configurations of the fields produced
classically by the mass M in the two configurations - the test particle, pulled
equally in both directions would not move at all!

       There are many prejudices against any such
theory in which only some of the
fields are quantised and the others are not. Such an approach would be an anti
-
thesis of the economy of thought that forms the basis of theoretical physics.
Moreover, in such theories it would be possible to have classically equivalent
theories - related by transformation of variables - to have inequivalent semi -
classical theories. A simple example explicitly demonstrating this is the
consideration of a gravitating conformally coupled
scalar field as a matter field, described by the action:
$$ S'[g',\phi'] = \int d^4 x \sqrt{-g'}[-R'/12\kappa^2 +
g'^{\mu\nu}\partial_\mu
\phi'\partial_\nu\phi' + R'\phi'^2/12]\eqno{(1.2)}$$
Under a matter field dependent redefinition of the metric
$$g'_{\mu\nu} = g_{\mu\nu}cosh^2\kappa\psi, \phi' = \kappa^{-1}tanh\kappa\psi$$
the action transforms to
$$ S[g,\psi] = \int d^4 x \sqrt{-g}[-R/12\kappa^2 + g^{\mu\nu}\partial_\mu
\psi\partial_\nu\psi]\eqno{(1.3)}$$
For both the actions one loop on - shell counter terms can be computed. These
must be added to the respective actions $S, S'$ in order to remove one loop UV
divergences in quantum theory. If the full action is quantised, then both S and
$S'$ give the same result for the counter terms: $\approx\int\sqrt{-g}R^2$.
However if only the the scalar field [and not the metric] is quantised, the one
loop divergences [which are exact as the scalar field in both cases suffers no
self interaction] give the counter terms as:
$$ \Delta S' \approx\int\sqrt{-g'}C'^2  = \int\sqrt{-g}C^2,
\Delta S\approx\int\sqrt{-g}[C^2 + 5R^2/3]$$
Thus on - shell counter terms illustrate the equivalence of $S$ and $S'$ if we
consider the full quantum theory but their in - equivalence if only matter
field is quantised. Therefore, unless one passes a drastic legislation
forbidding
matter dependent transformation of the metric, we have to consider a full
quantum theory. This means that one has to confront the divergences
that occur in gravitation. A simple dimensional analysis argument suggests that
if the coupling constant of a field is dimensional with dimension $m^d$ [in
units $c = \hbar = 1$], then the integral of a Feynman diagram of order N
behaves at large momenta like $\int p^{A - Nd}dp$, with A depending on the
physical process in question (number of external legs)
and not on d. Therefore for $d < 0$, integrals for
any process will diverge at a suffitiently high
order. For general relativity, the
dimension of the gravitational
constant is -2 [$G = 6.7\times10^{-39}Gev^{-2}$]. The
theory is thus dangerously non - renormalizable.

     A promising program, much exhaulted in the early 80's, was proposed by
Adler and Zee [Adler 1982]. The idea was to generate gravitation as an
effective theory.
Consider a renormalizable field theory with action:
$$ S[\phi^L,\phi^H] = \int d^4x L[\phi^L,\phi^H]$$
where the superscripts refer to the ``light" and the ``heavy" constituents
of the matter field. If the $\phi^H$ is not directly observable the functional
of the effective action of $\phi^L$ is expressible as:
$$Z = \int d[\phi^L]e^{iS_{eff}[\phi^L]}$$
where $S_{eff}[\phi^L] \equiv \int d[\phi^H]e^{iS[\phi^L,\phi^H]}$.
Accordingly, we
define
$$e^{iS_{eff}[g_{\mu\nu}]} = \int d[\phi]e^{iS_{eff}[\phi,g_{\mu\nu}]}
\eqno{(1.4)}$$
as the gravitational effective action induced by quantised (renormalised)
matter
fields on a curved spacetime background. This effective action is expressible
as:
$$ S_{eff}[g_{\mu\mu}] \equiv \int d^4x \sqrt{-g} L_{eff}[g_{\mu\nu}]$$
with $L_{eff}$ having a series expansion in powers of $\partial_\lambda
g_{\mu\nu}$
$$L_{eff}[g_{\mu\nu}] \equiv L_{eff}^{(0)}[g_{\mu\nu}] +
L_{eff}^{(2)}[g_{\mu\nu}]
+ O(\partial_\lambda g_{\mu\nu})^{(4)}$$
$$ \equiv {1\over 16\pi G_{ind}}(-2\Lambda_{ind}) + {1\over 16\pi G_{ind}}R
+ O(\partial_\lambda g_{\mu\nu})^{(4)} \eqno{(1.4)}$$
The following result, established by Adler, forms the basis of the program:

	Theorem: If there are no bare masses or massive regulators and if all spin
zero fields occur in massless super - multiplets, then calculable, induced,
$G_{ind}$ and $\Lambda_{ind}$ are produced by the above prescription.

	In addition if we chose to have a scalar field with a lagrange density
$L \sim \epsilon\phi^2R + T - V(\phi^2)$ where $V(\phi^2)$ may have a broken
symmetry phase with $\phi_{Min} = {\bar \phi}$, we get $[16\pi G_{ind}]^{-1} =
\epsilon {\bar \phi}^2$. In this case however, the scalar field and the
dimensional parameter ($m_0$ - the mass of the field) explicitly occur in the
theory. $G_{ind}$ is not
calculable in such a case. $\epsilon$ is an additional curved spacetime
parameter of the theory - not determined by the flat spacetime renormalised
parameters. In general if $S[\phi,g_{\mu\nu}]$ contains terms proportional to
R, then the finite renormalizations ambiguities arising from such terms would
produce an undetermined finite contribution to $G_{ind}$. The gravitational
constant would then be renormalizable but not calculable. We shall come
back to such an induction of a gravitational constant in more detail later.

	In a general renormalizable field theory we have a free renormalised
coupling or a mass parameter for each bare coupling or mass appearing in the
unrenormalised lagrange density. In quantum electrodynamics, for example, the
charge e is not calculable. To one loop order, the divergence in the charge
renormalization parameter $Z_e$ has the form:
$$Z_e = 1 + {\alpha_0\over 3\pi}logM^2 + O(\alpha_0^2), \alpha_0 = e_o^2/4\pi$$
M, being a massive regulator. Under a rescaling $M \longrightarrow \xi M$ of
the
regulator mass, $Z_e \longrightarrow Z_e + {\alpha_0\over 3\pi}log\xi^2$.
Thereby
implying that the finite part of $Z_e$ is regularization scheme dependent. Thus
the finite e extracted from the divergent bare charge $e_0$ remains a free
parameter in a renormalizable field theory. The result carries over
for higher order diagrams except that the renormalization constants associated
with the action density may no longer remain simple products of renormalization
constants of individual fields, charge and mass functions. The multiplicative
renormalization is taken over by more general form of matrix multiplication
renormalization. This in general mixes quantities in accordance with the
dimensional algorithm that allows for mixing of composite operators of the same
canonical dimension and symmetry type - with the lagrange density containing
a complete basis set of composite operators of canonical dimension 4.

	With the above constraints, the only way to get a calculable and non -
vanishing dimensional physical parameter like $G_{ind}$ is by the process of
dimensional transmutation
[see eg Stevenson 1981]. We have examples of theories which are scale
invariant at the classical level but exhibit spontaneous scale invariance
breaking as a result of quantum corrections in one or higher loop order. For
example consider an SU(n) non - abelian gauge field theory [without any scalar
fields] coupled to $N_f$ massless fermions in the fundamental representation.
At the classical level there are no dimensional parameters. When radiative
corrections are included, the coupling constant g appears in calculations
through a running coupling constant:
$$g^2(-q^2) = {g^2(\mu^2)\over 1 + g^2(\mu^2)log(-q^2/\mu^2)b_0 + ...}$$
[the dots ... representing the higher order terms]. q being the 4 - momentum,
$\mu^2$: an arbitrary subtraction point introduced because the massless
gauge theory is highly infra red divergent and $b_0 = [11n/3 - 2N_f/3]/8\pi^2$
(determined for one loop radiative corrections) is positive for $N_f$ not too
large. For $b_0 > 0$, the coupling constant goes to zero at large 4 - momentum
square - describing asymtotic freedom. Under a change of subtraction point
$\mu \longrightarrow\mu_1$ we get (to one loop):
$${1\over g^2(\mu_1^2)} - {1\over g^2(\mu^2)} = log[\mu_1^2/\mu^2]b_0/2$$
$$\Rightarrow {1\over g^2(\mu_1^2)} - log\mu_1^2b_0/2 = {1\over g^2(\mu^2)}
- log\mu^2b_0/2$$
$\Rightarrow M(g(\mu),\mu) \equiv \mu exp-[b_0g^2(\mu^2)$ is subtraction point
independent [also said to be renormalisation group invariant]. This result can
be generalised to all orders. Since all observables should be subtraction
point independent, they can depend on the scale $\mu$ only through the scale
invariant mass M - a circumstance captured by the theorem[Gross et al 1974]:
Any physical
parameter $P[g,\mu]$ of canonical dimension d must be equal to
$[M(g(\mu),\mu)]^d$
upto a calculable number. Thus instead of a one parameter family of
unrenormalised
theory characterised by the values of unrenormalised dimensionless couplings
$g_o$, the renormalisation process itself replaces it by a one parameter family
of renormalised theories characterised by the value of the dimensional scale
mass M. This process is refered to as dimensional transmutation. One could
adopt the strategy of equating the dimensional parameter to the physical
observable and identify its value by appealing to experiment. The
renormalisation
group equations then tell us that the value so fixed is invariant under the
change of the renormalisation scale.

	If we can realise a basic premise that the Einstein - Hilbert action is not a
fundamental action at all but an induced effect resulting from quantum
fluctuations of matter fields - we would save ourselves from the arduous task
of ``quantising gravity". In the context of what was just described,
$G_{ind}^{-1}$ and $\Lambda_{ind}$ induced in (say) a gauge field theory are
simply
physical parameters of canonical dimension 2. To extract expressions for these
parameters one considers deviations of the metric $g_{\mu\nu}$ from the
Minkowski
metric $\eta_{\mu\nu}$. Recalling the definition of $S_{eff}[g_{\mu\nu}]$
with $\phi$ a renormalised field in a curved spacetime background and
considering
a conformal variation around a general background metric by operating upon both
sides of eqn(1.4) by $2g_{\mu\nu}(y){\delta\over \delta g_{\mu\nu}(y)}$, gives:
$$2g_{\mu\nu}(y){\delta\over \delta g_{\mu\nu}(y)}\int d^4x\sqrt{-g}
[{1\over 16\pi G_{ind}}(-2\Lambda_{ind}) + {1\over 16\pi G_{ind}}R
+ O(\partial_\lambda g_{\mu\nu})^{(4)}]$$
$$ = {\int d[\phi]exp[iS[\phi,g_{\mu\nu}]]2g_{\mu\nu}(y){\delta\over
\delta g_{\mu\nu}(y)}\int d^4x\bar L\over \int d[\phi]exp[iS[\phi,g_{\mu\nu}]}
\eqno{(1.5)}$$
the quantities inside the x - integral being at the spacetime x. In terms of
standard definitions for the variations of the associated quantities:
$$ \delta\bar L \equiv {1\over 2}\bar T^{\mu\nu}\delta g_{\mu\nu},  \bar
T^{\mu\nu}
\equiv \sqrt{-g}T^{\mu\nu}$$
$$\bar T[g_{\mu\nu},x] \equiv \sqrt{-g}T^\mu_\mu;  T(x) \equiv
\bar T[\eta_{\mu\nu},x] = T^\mu_{\mu, g_{\mu\nu}=\eta_{\mu\nu}} \eqno{(1.6)}$$
[The field $\phi$ could even be the metric itself, with the associated
classical
action being a scale invariant [4th order] renormalizable action [Stelle
1977]].
Straight
forward formal manipulation gives:
$$ -{1\over 2\pi}{\Lambda_{ind}\over G_{ind}} = {\int
d[\phi]exp[iS[\phi,\eta_{\mu\nu}]]
T(0)\over \int d[\phi]exp[iS[\phi,\eta_{\mu\nu}]]}$$
$$ \equiv <T(0)>_0 \eqno{(1.7)}$$
Further, when the metric itself is not being quantised [the expression for the
case when the metric is also quantised can also be worked out[Adler 1982]], we
get the
value for the induced gravitational constant:
$${1\over 16\pi G_{ind}} = {-i\over 96}\int d^4x x^2<T^{*}(\tilde T(x)\tilde
T(0))>_0
\eqno{(1.8)}$$
Here $T^{*}$ represents the time ordered product, $\tilde T(x) \equiv T(x) -
<T(x)>$ and $x^2 = (x^0)^2 - (x^i)^2$. These flat spacetime formulae are to be
taken in their dimensional continuation limits:
$${1\over 16\pi G_{ind}} = {-i\over 96}Lim _{\omega\rightarrow 2}
\int d^{2\omega}x x^2<T^{*}(\tilde T(x)\tilde T(0))>_0^\omega
\eqno{(1.9)}$$
the vacuum expectation value in the integrand beign the value in the $2\omega$
dimensional spacetime. Denoting ``logs" as power series in $log(-x^2)$, the
vacuum expectation value in the integrand has an operator product expansion
$${<O_0>_0\over (-x^2)^4}\times logs + {<O_2>_0\over (-x^2)^3}\times logs
+ O[{1\over (-x^2)^2}]\eqno{(1.10)}$$
The $O_{0,2}$ are invariant operators of canonical dimension 0 and 2. The first
term in this expansion gives a quadratically divergent result vanishing
in the dimensional
continuation algorithm. The second term also gives a divergent result. The
third
term gives a finite result. Thus in a theory in which there is no $O_2$
present,
an effective Einstein - Hilbert action, with calculable gravitational and
cosmological constants, is dynamically induced.

	Explicit calculations, made for over a large class of renormalizable models
[Zee 1982],
realise the above prescription. The sign of the induced constants turn out
to be sensitive to the infra - red details of the renormalisation theory. There
is no general theorem regarding the sign of the induced constants.

Section II.	One requirement for the calculability of the induced constants is
the
absence of a matter - curvature coupling in the bare action. For a scalar field
exhibitting spontaneous symmetry breaking for example we noted that we could
have
an induced, non - calculable gravitational constant. We now explore the
possibility of having such couplings with an effective potential having an
absolute minimum at $\phi = 0$. Such a possibility may, in situations that
shall
be described below, not stand in the way of the calculability criteria for the
induced constants. This would be particularly realised if the scalar field in
question does not couple with the other renormalizable
matter fields at all. We find it appropriate to recount the history of the
scalar
field with reference to attempts made to consistently generate a classically
effective theory of gravity in the following.

 	The role of scalar field(s) to generate a dynamic theory of gravity [even at
the classical level] itself has invited a
considerable amount of attention. Typically in these approaches, the
gravitational action is effectively induced by a coupling of the scalar
curvature
with a function of a scalar field. The bulk properties of the universe fixes
the value of the function at a given epoch - thereby yielding an
effective Einstein - Hilbert action at that epoch. This is central to
the approaches of Scherrer [1950], Hoyle and Narlikar[1964], Brans $\&$ Dicke
[1961],
where
versions of Mach principle
motivated a judicious choice of the sign of coupling of a scalar field to the
curvature scalar to describe particular solutions in which the gravitational
constant gets related to the bulk properties of the Universe.

	Deser[1970] considered an $\omega = -3/2$ Brans - Dicke theory [which also
corresponds to a smeared out version of the Hoyle - Narlikar theory] described
by the conformally invariant action:
$$S[\phi, g_{\mu\nu}] = -\int d^4x\sqrt{-g}
[g^{\mu\nu}\partial_\mu\phi\partial_\nu\phi + R\phi^2/6] +
S_m[\psi]\eqno{(2.1)}$$
$S_m$ being the action for the other matter fields $\psi$.
Variation with respect to
the metric and the fields gives
$$\nabla^2\phi - {R\over 6}\phi = 0$$
$$G_{\mu\nu} = 6\phi^{-2}[T_{\mu\nu}[\phi] + T_{\mu\nu}[\psi]]$$
$$T_{\mu\nu}[\phi] \equiv \sqrt{-g}[\phi_\mu\phi_\nu
-{1\over 2}g_{\mu\nu}\phi_\alpha^\alpha
+ {1\over 6}(g_{\mu\nu}\nabla^2 - D_{\mu\nu})\phi^2] \eqno{(2.2)}$$
The trace of this equation requires $T^\alpha_\alpha[\psi] = 0$. Thus in such a
conformally invariant theory, the only matter fields that can consistently
(classically) couple are the scale invariant fields. Thus the photons would
weigh but the sun would not. To get a realistic model, one gives up conformal
invariance by introducing an extra term $\int d^4x\sqrt{-g}f[\phi]$. This gives
$$ [\phi f' - 4f] = 2T^\alpha_\alpha[\psi]$$
for the trace of the matter fields. For $f[\phi] = \mu^2\phi^2/2$,
$\mu^{-1}$ being
the range of the $\phi$ field, it would have to be chosen to be cosmological
in order not to be at variance with observations. Chosing it to be of the order
of ``the universe radius"
$\sim R_0, \Rightarrow T^\alpha_\alpha[\psi] = -\mu^2\phi^2$.
With the average of the matter trace $<T^\alpha_\alpha[\psi]>_{av} \sim
M_U/R_0^3$,
$M_U$ being the ``mass of the universe", gives an expression of a classically
induced ``gravitational constant":
$$G_{ind} \sim R_0^{-2}[{M_U\over R_0^3}]^{-1} \sim {R_0\over M_U}$$
This is a prototype of arguments used in such theories to relate the
gravitational constant to the bulk parameters of the universe.

			 In addition to the above, the incorporation
of spontaneous symmetry breaking in the picture leads to interesting
possibilities [Zee 1979, 1980]. Consider an action:

$$ S = \int d^4 x \sqrt{-g}[\epsilon\phi^2R/2 +
g^{\mu\nu}\phi_{,\mu}\phi_{,\nu}
- V(\phi) + L_w] \eqno{(2.3)}$$
where V is a potential minimised at some value $\phi = v, L_w$	being the
lagrangian for the other matter fields. The theory, at $\phi = v$, is
indistinguishable from Einstein's theory with the gravitational constant
$[G_N = (8\pi\epsilon v^2)^{-1}]$ dominated  - not by the bulk properties of
the universe but by the minimum point of V for small R. For large R - as in the
early Universe, the model would account for an adiabatic variation of the
gravitational constant. A substitution  $\phi = v + \psi$ requires the
existence
of a scalar particle $\psi$ with mass $V''(\phi = v)$ [Considering a potential
$\lambda (\phi^2 - v^2)^2/4$ with $\lambda < 1$, the mass of $\psi =
\lambda^{1/2}v
= (\lambda/8\pi\epsilon)^{1/2}M_{PLANK}$ which is of the order of $10^{19}$ Gev
for $\epsilon$ of the order unity. For a different choice of V one can have
smaller
values of the $\psi$ mass.]. As one goes back in time the scalar curvature
becomes larger, implying a variation $\delta G/G = -2\delta v/v = 2\epsilon
R/V''(v)$
$\longrightarrow$ (Hubble constant / mass of $\psi)^2$. This variation would
become
important when the age $[H^{-1}]$ of the universe were to be of the order of
the
Compton time of $\psi$. Thus in these theories, G is affected by the bulk
properties only at early times while at later times it is dominated just by the
minimum of the effective potential.

	Probing the possibility that the unification mass scale for weak, strong and
eletromagnetic interactions, as given by the SU(5) Georgi - Glashow model, may
be related to the plank scale itself, one may further consider  the
coupling of the scalar curvature to, in general, a set of Higg's fields
transforming under various representations of SU(5). In the early universe,
this would yield changes to $\delta G/G$ from rising temperatures. A naive
``restoration of symmetry at a sufficiently high temperature" would lead to the
effective gravitational constant becoming infinite at finite temperature.
However, one can quite consistently have no such
restoration effect [Weinberg(1974), Mohapatra et al(1979)].
For example
consider theory with two scalar fields $\phi \& \eta$ with the quartic part
of the potential $V_Q[\phi,\eta] = \lambda_1\phi^4 + \lambda_2\eta^4 +
2\lambda_3\phi^2\eta^2$. At high temperatures the potential gets modified
by addition of a term $(\lambda_1 + \lambda_3)T^2\phi^2$  (T being the
temperature).
Consistency with
positivity of the potential requires $\lambda_1 > 0, \lambda_2 > 0,
\lambda_1\lambda_2 > \lambda_3^2$. One can consistently have $\lambda_3 < 0;
\vert\lambda_3\vert > \lambda_1$. As a result a term $\sim -T^2\phi^2$ gets
added to the effective potential at high temperatures.
Restoration of symmetry at high temperatures would never take place in such
models. This leads to $\phi_{MIN}^2 \longrightarrow T^2$ at high temperatures.
The
effective gravitational constant weakens as $1/T^2$ in the early universe. For
the early universe, this implies that the conformal scale factor of the
Friedman - Walker metric goes linearly with time at such high temperatures.
Such a behaviour of the conformal scale factor leads to a simple resolution
to the horizon problem in cosmology [Rindler(1956)]:
the horizon radius, related to the
integral $\int dt[R(t)]^{-1}$ diverges for the conformal scale factor
$R(t) \sim t$ as the lower limit of integration goes to zero.

	In general scalar field theories, the presence of an $R\phi^2$ coupling is
necessary to ensure that the matrix elements of the physical energy momentum
tensor is independent of the cut - off in renormalised pertubation theory.
Consider the classical lagrangian:
$$L = {1\over 2}\partial_\mu\phi\partial^\mu\phi - V(\phi) - {R\over 12}\phi^2
\equiv \bar L - {R\over 12}\phi^2 \eqno{(2.4)}$$
As $R \sim (g_{\mu\nu,\mu\nu} - g_{\mu\mu,\nu\nu})$, the canonical stress
energy tensor gets extra contributions even in the limit of a flat spacetime:
$$\Theta_{\mu\nu} = \partial_\mu\phi\partial_\nu\phi - g_{\mu\nu}\bar L
- {1\over 6}(\partial_\mu\partial_\nu - g_{\mu\nu}\nabla^2 )\phi^2
\eqno{(2.5)}$$
with the equation of motion of the field being
$$ \nabla^2\phi + V'(\phi) = 0 \eqno{(2.6)}$$
we get
$$\Theta_\alpha^\alpha = 4V(\phi) - \phi V'(\phi) \eqno{(2.7)}$$
as the on - shell expression for the trace. In the absence of the $R\phi^2$
coupling, the trace of the conventional stress energy tensor would involve
the derivative of the field as well. This leads to difficulties.
We recall some standard results in the BPH regularisation scheme [Hepp(1965),
Callen
et al (1970)]:

	Consider a Lagrangian that is polynomial in fields and derivatives, and
consider the Green's functions in pertubation theory. At some stage in the
expansion, one - particle - irreducible diagrams with a superficial degree
of (cut - off dependent) divergence d appear. One then adds counter terms to
the lagrangian involving as many fields as the external lines of the diagram
and upto d derivatives. Choosing the coefficients of these counter terms
to cancell the first d terms in the Taylor expansion of the divergent diagrams
about the point at which all the external momenta vanish, gives a resultant
expansion which has a finite limit in every order. For an ordinary
renormalizable theory, the counter terms have the same form as the original
terms in the lagrangian. The total lagrangian is thus of the same form as the
original. Rescaling the fields so that the kinetic term is of the standard
form,
the coefficients of the other terms can be identified as bare terms [masses,
coupling constants etc.]. In ither words, the Green's functions for rescaled
fields can be made cut - off independent by chosing the bare parameters in
a cut - off dependent way.

		Consider a change in any renormalisable field theory due
additional terms which are a combination of a set of monomials $A_i$
in the fields and their derivatives:
$$L \longrightarrow L + \sum A_ij_i \eqno{(2.8)}$$
the $j_i$ being arbitrary functions of spacetime. The $\Gamma_i^{(n)}$ may now
be defined as the Fourier transform of the variational derivative of the n -
point Green's function in the fundamental fields with respect to $j_i$, at
$j_i = 0$. The above considerations imply that the $\Gamma$'s can be made
finite
if appropriate counter terms are added to the lagrangian. If the counter terms
are also in the set $A_i$, the set is said to be closed under renormalisation.
Thus in such a case , cut - off independent functions $R_i^{(n)}$ can be found
such that the $\Gamma$'s are linear combinations of these R's. If one applies
these considerations to the conventional stress energy tensor, it turns out
that
a large number of terms are required to be added to
form a set of operators closed under renormalisation and too few Ward
identities
to determine the coefficients of the $R_i$'s. This problem does not arise for
the stress tensor $\Theta_{\mu\nu}$ arising upon the inclusion of the $R\phi^2$
term[Callan (1970)].

			From the above discussion we are motivated to consider
the effect of a renormalisable
scalar field theory in curved spacetime on the one hand, and a set of other
renormalised matter fields on the other - with the absence of any coupling
between the matter fields and the scalar fields [in the simplest case]. The
effective classical action is described by:
$$S = \int d^4 x \sqrt{-g}[-{\epsilon\over 2}\phi^2R +
{1\over 2}g^{\mu\nu}\phi_{,\mu}\phi_{,\nu}
- V(\phi) + \beta_{ind}R + \Lambda_{ind} + L_w] \eqno{(2.9)}$$
where all expressions in the action refer to renormalised [classically
effective]
quantities, $L_w$
being the action for the rest of the matter fields. We shall now see that in
such
an effective theory, classical, non - topological soliton solutions exist.
These
would lead to measurable spatial variation of the effective gravitational and
cosmological constants.

	Section	III. Solitons
[stable, fnite energy, non - trivial, classical solutions]
have been studied in a large family of field theories in four spacetime
dimensions in particular [see eg. Rajaraman 1982].
For a simple multi - component scalar field
theory with a potential which is positive definite throughout, Derrick's
theorem establishes the non - existence of any non - trivial, stable, static
solutions in three or more dimensions. In fact for potentials that are
positive definite everywhere, there are no solutions (stable or unstable).
However,
it is possible to get around Derrick's result. Coleman [1985]
demonstrated that for classical theories having a conserved charge
Q, associated with an unbroken symmetry of the theory, stable solutions which
are time dependent in the internal space and spherically symmetric in the real
space can exist. For example, consider the potential $U(\phi)$
for an SO(2) invariant theory
involving two real scalar fields $(\phi_1,\phi_2)$, positive everywhere, but
having more than one local minima as described in Figure [1]. The SO(2)
invariance
preserving time dependence
$\phi_1 = \phi (r)cos \omega t, \phi_2 = \phi (r)sin\omega t$ reduces the
classical equations of motion to:

$$ \phi'' + 2\phi'/r + \omega^2\phi - U'(\phi) = 0 \eqno{(3.1)}$$

Thus looking for a time dependent solution is equivalent to the search for a
spherically symmetric time independent solution in a theory with an effective
potential $ - {1\over 2}\omega^2\phi^2 + U$ [figure (2)] violating
the positivity condition
of Derrick's theorem. Coleman demonstrated that for fixed charge, non trivial
solutions do exist and are stable. Defining:
$$ \mu^2 = U''(0) = [2U/\phi]_{\phi = 0}$$
$$ min[2U/\phi^2] = 2U_o/\phi_o^2 = \omega_o^2 \eqno{(3.2)}$$
stable solutions exist for $\omega_o^2 < \mu^2$. The solution is represented by
a monotonically decreasing function $\phi(r)$ which goes to a constant
$\phi = \phi_o$ inside a radius $ R_b$. Outside the radius $\phi$ goes to zero
and the two regions are separated by a transition zone of thickness of the
order
of $\mu^{-1}$. The radius $R_b$ is related to the conserved charge. For large
Q: $Q = 4\pi R_b^3\omega_o\phi_o^2$. Upon an adiabatic alteration of the
parameters of the theory: as the minimun of U [i.e. $U_+$]
goes to zero, $\phi_o$ goes to
$\phi_+, \omega_o$ goes to zero and $R_b$, for fixed Q becomes large - going
to infinity in the limit.

	These results carry over to a curved spacetime with a small scalar curvature
coupling to $\phi^2$ in the class of theories described by the
action eqn(2.9), with $\phi$ replaced, in general, by a multi - component
field with a coupling $\epsilon\phi^2R$.
It would give rise to stable ``balls of gravity" [henceforth called
g - balls] even if we are not at [but approaching] a first order phase
transition. The effective gravitational
constant would be given by:

$$ G_{Eff} = G_I/[1 + \epsilon G_I<\phi^2>] \eqno{(3.3)}$$
Outside the g - ball, the dynamics of gravitation would be governed by the
induced Einstein - Hilbert action with the gravitational constant given by
$G_I = [8\pi\beta_I]^{-1}$. The effective gravitational constant inside the
ball would be given by $[16\pi(\epsilon\phi_o^2 + \beta)]^{-1}$
and the cosmological constant by $-U_+(\phi_+) + \Lambda_I$.

	Thus if one assumes that the stability argument for a Q - ball would
remain valid for curved spacetime, at least for small curvature, then on
account of the non - minimal coupling, this gives rise to a soliton with a
different effective gravitational constant inside that its corresponding
value outside.

	The existence of such solitons for non - minimally coupled theories
need not necessarily require a conserved charge. Indeed, the existence of
g - balls is also suggested for a single
component
scalar field in curved spacetime. Consider the variation of eqn(2.9) with
respect
to the background metric $g_{\mu\nu}$:

$$ [R_{\mu\nu} - g_{\mu\nu}R/2](\epsilon\phi^2/ + 2\beta_I)
+ (\epsilon\phi^2)_{;\mu;\nu} - g_{\mu\nu}(\epsilon\phi^2)^{;\rho}_{;\rho}
+ \phi_{,\mu}\phi_{,\nu} +$$
$$- g_{\mu\nu}[{1\over 2}\phi_{,\rho}\phi^{,\rho} - V(\phi)]
= - T_{\mu\nu}^w \eqno{(3.4)}$$
Here $T_{\mu\nu}^w$ is the stress tensor of the other matter fields in $L_w$
and
we have absorbed $2\Lambda_I$ in a redefinition of $V(\phi)$.
Variation with respect to $\phi$ gives:

$$ \phi^{,\rho}_{;\rho} - \epsilon R\phi + V'(\phi) = 0 \eqno{(3.5)}$$
R can be eliminated from this equation and the trace of eqn(3.4) to get an
equation of the form:

$$ \phi^{,\rho}_{;\rho} + W'(\phi) = 0 \eqno{(3.6)}$$

with

$$ W'(\phi)[2\beta_I + \phi^2\epsilon + 6\epsilon^2\phi^2] =
(2\beta_I +\phi^2\epsilon)V'(\phi)$$
$$ - \epsilon\phi[4V + T^{w\alpha}_\alpha - (1 + 6\epsilon)\phi^\rho\phi_\rho]
\eqno{(3.7)}$$
One could look for static spherically symmetric solutions to this equation for
$-W(\phi)$ having the profile given in figure(2) for a suitable choice of
parameters defining V. We express the general spherically symmetric metric
as:
$$ds^2 = e^{\nu(r)}dt^2 - e^{\lambda(r)}dr^2
- r^2[d\theta^2 + sin^2\theta d\phi^2] \eqno{(3.8)}$$
For vanishing $T^w_{\mu\nu}$, eqns. (3.4) and (3.6) reduce to the following
set:
$$(2\beta + \epsilon\phi^2)[-{1\over r^2} + e^{-\lambda}({1\over r^2}
- {\lambda'\over r})] = - V(\phi) - {1\over 2}e^{-\lambda}(\phi_{,r})^2$$
$$ -\epsilon e^{-\lambda}[(\phi^2)_{,rr} + (\phi^2)_{,r}({2\over r}
- {\lambda'\over 2})] \eqno{(3.9a)}$$
$$(2\beta + \epsilon\phi^2)[-{1\over r^2} + e^{-\lambda}({1\over r^2}
+ {\nu'\over r})] = - V(\phi) + {1\over 2}e^{-\lambda}(\phi_{,r})^2$$
$$ -\epsilon e^{-\lambda}(\phi^2)_{,r}({2\over r}+ {\nu'\over 2})
\eqno{(3.9b)}$$
$$(2\beta + \epsilon\phi^2)[{1\over r^2} - e^{-\lambda}({1\over r^2}
- {(\lambda' - \nu')\over 2r})] = + V(\phi) $$
$$ -\epsilon e^{-\lambda}[{-1\over r}(\phi^2)_{,r} -{1\over 2}(\phi^2)_{,rr}
- {1\over 2}(\phi^2)_{,r}({2\over r} - {\lambda'\over 2} +{\nu'\over 2})]
\eqno{(3.9c)}$$
$$ -e^\lambda\phi_{,rr} -e^\lambda\phi_{,r}[{\nu'-\lambda'\over 2} +{2\over r}]
+ W'(\phi) = 0 \eqno{(3.10)}$$
Defining a ``mass function'' $m(r)$ by $e^{-\lambda}\equiv 1 - 2m(r)/r$, one
gets:
$$2m'(r)(2\beta + \epsilon\phi^2) \approx r^2[V + e^{-\lambda}(W +
\epsilon[2\phi W' + 4W])] \eqno{(3.11)}$$
We look for solutions having $\phi$ near the (lower) minimum of $W(\phi)$ for a
suffitiently large $r [0 \leq r \leq R_b]$ and thereafter making a transition
to $\phi= 0$ over a further small distance $\Delta R$, and further thereafter
staying at $\phi = 0$ [the higher minimum of $W(\phi)$]. For $r \leq R_b$,
$e^{-\lambda} = 1 + r^2/\Lambda_{in}^2$ where $\Lambda_{in}^{-2} \equiv
- 8\pi G_{in}\rho/3 \equiv -\rho[\epsilon\phi_o^2 + 2\beta]^{-1}$
and we have defined
$V(\phi_o) \equiv \rho$. The thin wall approximation, appropriately defined as
the
condition that the second term of eqn(3.10) be negligible in comparison to the
other terms. This is equivalent to having $R_b$ and $\Lambda$ large in
comparison
to the domain of variation of $\phi$ [Coleman et al 1980]. In this
approximation,
eqn(3.11) can be integrated across the surface to read:
$$\Delta m(r) \approx R_b^2 \int_o^{\phi_o}d\phi [V + W(1+4\epsilon)
+ 2\phi\epsilon W'](2W)^{-1/2}/2(2\beta + \epsilon\phi^2) \eqno{(3.12)}$$
In the same approximation, given any potential $V(\phi)$ which gives rise
to an effective $W(\phi)$ having a double hump minima $|W(\phi) - W(0)|$ small,
it is in principal possible to integrate over the eqns (3.9) and (3.10).
Ideally one would like to start with a bare scalar field theory, take care
of all quantum corrections on account of self interaction of the scalar
field by integrating over all loops to all orders in pertubation theory to get
an effective potential. A non trivial classical solution which minimizes the
energy for the effective potential would then be both classical stable as well
as stable against bare quantum fluctuations [the same having been taken into
account in the resultant effective field theory]. Unfortunately, the effective
potential to all orders in pertubation has not been worked out. For a large
class of bare scalar field potentials, it was hoped that some general
convexity theorem could be used to establish essential characteristics of the
effective potential [Illiopolous et al, 1975]. However, more recent work has
cast
doubts on such hopes [Sher 1989]. We shall therefore not start with a guess
of an effective potential and integrate eqns(3.9) and (3.10). For our
purpose, we shall establish the stability of such a solution which follows
from the qualitative behaviour of the surface and internal energy
expressions.

	For a slowly varying $e^{-\lambda} \approx 1$, eqn(3.10) implies
that the distance over which $\phi$ has an appreciable variation from its
constant values [inside ond outside the ball of radius $R_b$] is of the
order of $ \Delta R \approx [W''(0)]^{-1/2}$. We shall assume this surface
thickness to be small as compared to $R_b$ - the radius of the ball. Inside
the ball, the expression for total energy is given by the volume integral
times the constant density $V(\phi) = \rho$:
$$E_{INT} = 4\pi\rho\int_o^{R_b}{r^2dr\over \sqrt{1+r^2/\Lambda^2}}$$
$$ = 2\pi\rho\Lambda^3[{R_b\over \Lambda}\sqrt{1+{R_b^2\over \Lambda^2}}
- ln({R_b\over \lambda} + \sqrt{1+{R_b^2\over \Lambda^2}})] \eqno{(3.13)}$$
The only property of the effective potential $V(\phi)$ and the associated
$W(\phi)$ that we have used so far is the profile of $W(\phi)$ given in
the figure[2] namely that $|W(\phi_o) - W(0)|$ is a small quantity. The
contribution to the surface energy would imply an additional geometric mass
given by eqn(3.12). We choose the potential $V(\phi)$ and its associated
$W(\phi)$ to be such that this equation is integrable. In particular we
assume this to be possible even if $(2\beta + \epsilon\phi^2)$ were to
change sign in the region $ R_b \leq r \leq R_b + \Delta R$. In general
therefore, the surface energy would be of the form $E_S = 4\pi R_b^2A$ with
$A$ depending on an appropriate integral across the surface (if necessary)
defined in terms of the principle value of the geometric mass.

	To obtain the total energy of the ball, one must add the gravitational
energy. This is the energy required to assemble the ball of radius
$R_b+\Delta R$ of constant density $\rho$ inside radius $R_b$ and surface
energy $4\pi R_b^2A$, on account of the gravitational potential determined by
the external gravitational constant $G_o$. Consider a sphere of
radius $r$, density
$\rho$ having the gravitational potential $V_G = -4\pi G_or^2\rho/3$.
Increasing
the  thickness by $dr$ by bringing an additional mass $\rho 4\pi r^2dr$ from
$\infty$ to $r$ requires energy $dE(r) = -16\pi^2r^4dr/3$. This is to be
integrated for $r$ going from zero to $R_b$ to get
$E_G = -{16\over 15}\pi^2G_o\rho^2R_b^5$. Further the work done to bring the
surface of mass $4\pi R_b^2A$ from $\infty$
to the surface is $-16\pi^2G_o\rho AR_b^4$.
The total energy is thus:
$$ E_{Total} = E_{INT} + 4\pi R_b^2A + E_G - 16\pi^2G_o\rho AR_b^4
\eqno{(3.14)}$$
{}From this it is clear that for suffitiently small $\rho$, it is possible
to have
$${dE_{Total}\over dR_b} = 0; {d^2E_{Total}\over dR_b^2} > 0$$
The value of $R_b$ for which this happens defines the gravity ball. If the
expression eqn(3.14) does not vanish for any value of $R_b$, it signals the
stability of the vacuum [Coleman et al 1980]. For example, with A > 0 and
for $E_{Total} > 0$ for $R_b$ satisfying eqn(3.15). $E_{Total}$ does not cross
zero for any value of $R_b$ and the solution is then stable against the decay
of the exterior of the ball to a true vacuum state which has $\phi = \phi_O$
everywhere.

	III. {\bf Discussion}: What we have explored is the possibility that
both dynamical as well as spontaneous symmetry breakdown play an important
role in getting an effective theory of gravitation. In such a program one
avoids
the difficulties encountered in the quantisation program of the Einstein -
Hilbert theory. The Einstein - Hilbert theory is thus induced as an effective
theory on account of a dynamical breakdown of scale invariance by standard
prescription of dimentional transmutation. Further if one has, in addition,
non minimally coupled scalar fields [with an effective potential that can
exhibit spontaneous symmetry breaking, one can have solitons in the theory.
Such regions would appear as gravity balls - with an effective gravitational
and
cosmological ``constants" differing in their values on the inside and outside.
In contrast to earlier proposals of involving scalar fields to dynamically
generate a theory of gravitation [eg. the theory of Brans and Dicke theory
which, in its original form requires a very high value of the
coupling parameter to be consistent with observations [Weinberg 1982]], the
spontaneous breakdown of symmetry induced by a suitable choice of an
effective potential of the scalar field, ensure the existence of gravity balls.
For a multicomponent scalar field having an associated conserved charge,
such solutions would occur even if we are not at but approaching a first order
phase transition. The radius of such balls would depend on the total charge
and also (weakly) on the trace of the rest of the matter's stress tensor.

	We have considered several possible applications of the effective
theory outlined in this article. If g - balls are small [$\approx 10^2$ cms],
they would be uninteresting. The observed gravitational constant would then
have to be fitted with the value external to the ball. One can envisage such
balls to contribute to dark matter. Larger ball [$\approx 10 - 10^2$ mrts.]
would perturb a Cavendish  experiment if it engulfs the apparatus. Balls of
a much larger size : terrestial or the size of a typical star, would completely
destabilize a gravitationally bound system as the same crosses the surface of
the ball. A ball of a size of a typical gallactic cluster would have an
interesting bearing on the dark matter problem at the scale of a gallactic
cluster. It is tempting to be inside such a ball i.e. have the effective
constant
inside to have the Newtonian value. Outside the g - ball, the spacetime
dynamics
would be governed by
the induced Einstein - Hilbert action. Observations are consistent with a very
small $\Lambda_I < 10^{-54} cm^{-2}.$ However one has no independent
measure of $\beta_I$.
The universality of the gravitational constant is assumed while determining the
large scale characteristics of the Universe. The effective gravitational
constant inside the ball would be given by $[16\pi(\epsilon\phi_o^2 +
\beta)]^{-1}$
and the cosmological constant by $-U_+(\phi_+) + \Lambda_I$.

	It is very tempting to choose $\Lambda_I$ to be vanishingly small while
holding $U_+$ to a small value. This would imply a negative cosmological
constant inside the g - ball. An appropriate value of $-\Lambda_{Eff}$ (of the
order of $10^{-51} cm^{-2}$) would fit velocity rotation curves without
interfering with
smaller (solar system) scale gravitational predictions. The velocity curve, for
such a value of $\Lambda_{Eff}$, for stars in a typical galaxy with a
maximum edge of the core speed given by $v/c$ of the order of $10^3$ is
shown in figure [3]. It is also quite tempting
to consider a value of $\beta_I$ smaller, than its Newtonian value inside the
g - ball, by a whole order of magnitude. This would prop - up the estimates of
masses of far - away objects obtained from virial speeds.

	 In general one could
consider both signs of $\beta_I$ (and hence $G_I$
and $\Lambda_I$. The dependence of the g - ball radius on the trace of the
stress
tensor would also lead to interesting effects. Given a value of the trace
$T^w$, one
could evaluate the value of the radius of the soliton. If the radius of the
soliton is greater than the extent of a chunk of matter distribution having the
trace $T^w$, the characteristics of the ball are not going to change
appreciably.
However, for a large distribution of matter and for a critical $T^w$ such that
$|W(\phi_o) - W(0)| \longrightarrow 0$, [or in the case of a Q - ball of scalar
matter, non - minimally coupled to curvature, as $U_+ \longrightarrow 0$ for
fixed Q], the radius of the g - ball can be made arbitrarily large. We have
explored several applications of such a scnario to earliy universe cosmology
and we describe an interesting ``toy cosmological model''.

	Consider a Freidman - Robertson - Walker cosmology with a difference: having
the gravitational constant with a ``wrong'' sign: $G_I < 0$. From the ensuing
equations:
$$ ds^2 = R(t)^2 [dr^2/(1 -kr^2) + r^2[d\theta^2 + sin^2\theta d\phi^2] - dt^2
\eqno{(4.1)}
$$
the dynamics is governed solely by the R term. For cosmic dust of co - moving
density $\rho$, the vanishing of the covariant divergence of the stress tensor
yields:
$$ \beta\rho R(t)^3 = constant = M_0 (say)$$
The Friedman equation reads:
$$R'(t)^2 = \Lambda_IR(t)^2/3 - k - M_0/R(t) = F[R(t)] \eqno{(4.2)}$$
It is easy to see that for $\Lambda_I \approx 0$, the only possible solution
is the $k = -1$ solution. Further, for arbitrary values of the gravitational
constant and the density parameter $\rho$, the solution quickly approaches the
Milne universe solution $R(t)^2 \longrightarrow 1$. The solution has no past
horizon and thus there is no horizon problem as the solution to eqn(4.2)
concaves
to the Milne limit. There is no initial singularity at all as every co moving
volume emerges from a non - vanishing radius in the past [$R(t)$ never
vanishes].
Not only would such a universe never enter the ``quantum gravity era'', there
is
no problem in the quantum graavity program - as by construction, the Einstein -
Hilbert theory is merely an induced theory arising from a well behaved quantum
program of an appropriate renormalisable field theory. With a negative
gravitational constant, a homogenous, isotropic metric is no longer a very
special metric [as is the case in canonical cosmology]. The story of such a
universe can be picked up from the epoch when any co - moving volume has a
minimum
radius [vanishing of the right hand side of eqn(4.2)]. We only require all
matter
to be highly relativistic at that epoch. As the universe expands and cools,
some
matter starts becoming non relativistic and the stress tensor starts acquiring
a
non - vanishing trace which rises to a maximum and then falls steadily to its
present value determined by the low average density at the present epoch. At
some
stage we shall encounter with gravity balls which we fix [by an appropriate
choice of the effective potential] to be such that $G_{inside} = G_{Newton}$.
The non - vanishing trace would determine the radius of the gravity ball. We
assume that at some epoch the value of the trace $T^w$ is such that
$|W(\phi_o) - W(0)| \longrightarrow 0$. A large gravity ball would now be
formed
within which we may accomodate the normal scenarion of light element systhesis.
As matter continues to expand, the $W(\phi_o), W(o)$ difference would
destabilise large balls. However, smaller, stable balls would exist.
We have not been able
to formulate a mechanism of bifurcation of unstable balls - however, if we
assume this to be possible, one can visualize  a top - bottom heirarchy of
structure formation. The smallest gravity ball being of a typical cluster size
in the model. All above speculation would have to be restrained to an
exploratory spirit. This is because we do not have a formulation that would
enable us to calculate the rate of formation of such balls - their coalescence
or fragmentation - to make a more
fruitful contact with large scale structure in
Cosmology.

	We may also tempt ourselves to the possibility of g - balls with
$G_{out} = G_{Newton}$ and $G_{in}$ to be much larger and attractive. The
typical
size of such balls being of the order of say $10^3$ to $10^4$ astronomical
units.
The interior of the ball may further have an effective [negative] cosmological
constant. The interaction of a $10^2$ to $10^3 M_\odot$ cloud entering the
ball,
as observed from far in the external region is what we are exploring as a
candidate for a model of a quasi - stellar object [without having to push it
to cosmological distances]. The high $G_{in}$ would cause a much more rapid
evolution of the section of the cloud that is sucked inside the ball. The
evolution can be expected to lead to very rapid - high metalicity of the
collapsed section. Further, a large negative $\Lambda_{in}$ would contribute
to the red shift. A preliminary study has revealed that we can account for
several qualitative features of the spectrum emmitted by quasars. Details
are being prepared for a separate report.

\vskip 1cm

Acknowledgement: The work presented in this article was developed during the
author's visit to imperial College London and DAMTP Cambridge in 1992 - 93
and portions of it were presented in seminars in the joint DAMTP - Imperial
College Quantum Gravity workshops.
Heplful discussions with Prof. T.W. Kibble, Dr. Gary Gibbons, Dr. R. J. Rivers
and
Dr. M. D. Pollock are gratefully acknowledged. The work was supported by the
EEC fellowship grant.

\vfil\eject

\centerline{\bf References}

  Abbot L.F., Deser S., Nucl. Phys. B195, 76 (1982)

  Adler S.L., Rev. Mod. Phys. 54 (1982) 729

  Brans C., Dicke R., Phys. Rev. 124, 925(1961); Phys. Rev. 125, 2163 (1962).

  Callan C., Coleman S., Jackiw R., Ann. Phys. (NY) 59, 42  (1970)

  Coleman S., Nucl. Phys. B262, 263 (1985)

  Coleman S. and Deluccia F., Phys Rev D21, 3305 (1980)

  Derrick G.H., J. Math. Phys.,5, 1252 (1964)

  Deser S., Ann. Phys. 59 (1970) 248

  Gross D.J., Neveu A., Phys. Rev. D10, 3235 (1974).

  Hepp K., Comm. Math. Phys., 1, 95 (1965)

  Hoyle F. and Narlikar J. V., Proc. Roy. Soc. (Lond) A282, (1964) 190

  Illiopolous J, Itzykson C, Martin A, Rev. Mod. Phys. 47, 165 (1975)

  Kibble T.W. in Quantum Gravity, Oxford symposium, 1982.

  MacCallum, M.A.H., in ``General Relativity'' - an Einstein Centenary
volume, eds. Hawking, S.W., Israel W. (1979)

  Mohapatra R.N., Senjanovic G., Phys. Rev. Lett. 42, 1651 (1979)

  Padmanabhan,T. ``Cosmology today - Models and Constraints'', VI
IAU meeting on Astronomy (1994).

  Rajaraman R., ``Solitons and instantons", Northolland, 1982.

  Rindler W., Mon. Not. Roy. Ast. Soc. 116, 663 (1956)

  Scherrer W. Helv. Phys. Acta 23, 547 (1950)

  Sher M., Phys. Rep. 179, 273 (1989)

  Stelle K.S., Phys. Rev. D16, 953 (1977).

  Stevenson P.M., Ann. Phys. (NY), 132, 383 (1981).

  Weinberg S., Phys. Rev. D9, 3357 (1974)., ``Gravitation and Cosmology"
(Wiley, New York, 1982).

  Zee A., in ``Unity of forces in the Universe" Vol II, Ed. A. Zee,
P. 1062, World Scientific (1982). Phys. Rev. Lett. 42, 417 (1979);
Phys. Rev. D23, 858 (198 ); Phys. Rev. Lett 44, 703 (1980).

\vskip 2cm

Figure Captions:

[A]  ``Figure 1"

[B]  ``Figure 2"

[C]  ``Figure 3"

\bye